\documentstyle[prl,aps]{revtex}

\begin{document}
\title{General Strategies for Discrimination of Quantum States}
\author{Chuan-Wei Zhang, Chuan-Feng Li, Guang-Can Guo\thanks{%
Email: gcguo@ustc.edu.cn}}
\address{Department of Physics and Laboratory of Quantum Communication and Quantum\\
Computation,\\
University of Science and Technology of China,\\
Hefei 230026, People's Republic of China\vspace*{0.35in}}
\maketitle

\begin{abstract}
\baselineskip24pt We derive general discrimination of quantum states chosen
from a certain set, given initial $M$ copies of each state, and obtain the
matrix inequality, which describe the bound between the maximum probability
of correctly determining and that of error. The former works are special
cases of our results.

\newpage\ 
\end{abstract}

\baselineskip24pt It is determined by the principle of quantum mechanics [1]
that we can't perfectly discriminate or clone an arbitrary,unknown quantum
states. But this important result doesn't prohibit discrimination and
cloning strategies which has a limited success. Numerous projects has been
made to this subject by many author. Discrimination of states has a close
connection with quantum measurement. The general approach of quantum
measurement theory has been consider a quantum system whose unknown state
belongs to a finite, known set and devise the measurement which yields the
most information about initial state, where the figure of merit is the
probability of a correct result or the mutual information [2]. We have known
that a set of orthogonal quantum states can be discriminated perfectly.
Approximate state determination of a set of non-orthogonal states is also
possible. Helstrom has found the absolute maximum probability of
discriminating between two states, which is instead given by the well-known
Helstrom limit [2]. Non-orthogonal quantum states can also,with some
probability, be discriminated without approximation. Ivanovic [3], Dieks [4]
and Peres [5] have showed that it is possible to discriminate exactly ,which
means with zero error probability, between a pair of non-orthogonal states
and they derive the maximum probability of success called IDP limit. The IDP
limit is not the absolute maximum of the discrimination probability, but is
rather the maximum subject to the constraint that the measurement never give
error results. Chelfes and Barnett [6-10] extent their results to the
constraint of $M$ initial copies of states and also they consider the
results for $n$ states. They found that $n$ non-orthogonal states can be
probabilistic discriminate without error if and only if they are linearly
independent. Their works connect the approximate and exact discrimination
and give an inequality that describe the relation between the probability of
correctly and error discriminating. All their works have use an assumption
that the probabilities of correct or error for discrimination are equal for
different state. More recently, Duan and Guo [11] have found the maximum
probabilities of exact discrimination of $n$ linearly independent quantum
states and for different state the maximum probability may be different.
Exact discrimination attempts may give inconclusive results,but we can
always know with certainty whether or not the discrimination has been a
success. Recently, however, Massar and Popescu [12] and Derka et al [13]
have considered the problem of estimating a completely unknown quantum
state, given $M$ independent realizations.

In quantum mechanics, a combination of unitary evolution together with
measurements often yields interesting results, such as the quantum
programming [14], the purification of entanglement [15],and the
teleportation [16] and preparation [17] of quantum states. Duan and Guo [11]
used such combination in the field of quantum cloning and showed that the
states secretly chosen from a certain set $S=\left\{ \left| \Psi
_1\right\rangle ,\left| \Psi _2\right\rangle ,...,\left| \Psi
_n\right\rangle \right\} $ can be probabilistically cloned if and only if $%
\left| \Psi _1\right\rangle $,$\left| \Psi _2\right\rangle $,$...$,and $%
\left| \Psi _n\right\rangle $ are linearly independent. They derive the best
cloning efficiencies and also extent their results to the discrimination of
a set of states and give the optimal efficiencies for exactly
probabilistically discriminating. In this paper we'll also use such
combination and construct general discrimination of the states secretly
chosen from a certain set $S$, given $M$ initial copies of each state. We
derive the matrix inequality, which describe the bound between the maximum
probability of correctly determining and that of error. We give the most
general results about quantum discrimination in a finite set of states and
prove that the former works can be derived from our results in different
conditions.

Consider a set of quantum states $S=\left\{ \left| \Psi _1\right\rangle
,\left| \Psi _2\right\rangle ,...,\left| \Psi _n\right\rangle \right\} $ ,$%
S\subset {\cal H}$ ,where ${\cal H}$ is a $n$-dimension Hilbert space. If
there are $M$ quantum systems, all of which are prepared in the same states,
then we can denote the possible states of the combined system are the $M$%
-fold tensor products 
\begin{equation}
\left| \Psi _i\right\rangle ^{\otimes M}=\left| \Psi _i\right\rangle
_1\otimes ...\otimes \left| \Psi _i\right\rangle _M\text{ }
\end{equation}

Any operation in quantum mechanics can be represented by a unitary evolution
together with a measurement. We use a unitary evolution $\hat U$ and yield

\begin{equation}
\hat U\left| \Psi _i\right\rangle _A^{\otimes M}\left| P_0\right\rangle =%
\sqrt{\gamma _i}\left| \varphi _i\right\rangle _A\left| P_0\right\rangle
+\sum_jt_{ij}\left| \varphi _j\right\rangle _A\left| P_0\right\rangle
+\sum_jc_{ij}\left| \varphi _j\right\rangle _A\left| P_1\right\rangle ,
\end{equation}
where $t_{ii}=0$, $\gamma _i\geq 0$, $t_{ij}\geq 0$, $\left\{ \left| \varphi
_i\right\rangle _A,i=1,2,...,n\right\} $ are a set of orthogonal states in
Hilbert space ${\cal H}^{\otimes M}$, $\left\{ \left| P_0\right\rangle
,\left| P_1\right\rangle \right\} $ are the orthogonal basis states of the
probe system $P$, and $A$ represent the system of initial states. After the
unitary evolution we measure the probe system $P$. If we get $\left|
P_1\right\rangle $,the discrimination fails and we discard the output, so we
call this situation as an inconclusive result, else if we get $\left|
P_0\right\rangle $, we then measure system $A$ and if get $\left| \varphi
_k\right\rangle _A$ ,we determine the initial states as $\left| \Psi
_k\right\rangle ^{\otimes M}$. We can see the discrimination may be
correctly success (when $k=i$, the probability is $P_D^{(i)}=\gamma _i$) or
have errors (when $k\neq i$, the all probability is $P_E^{(i)}=%
\sum_kt_{ik}^2 $) when the initial state is $\left| \Psi _i\right\rangle
^{\otimes M}$.We can also give the inconclusive probability $P_I^{\left(
i\right) }=1-P_D^{(i)}-P_E^{(i)}$.

Unitary evolution $\hat U$ exist if and only if [18]

\begin{equation}
X^{(M)}=(\sqrt{\Gamma }+T\ )(\sqrt{\Gamma }+T\ ^{+})+CC^{+},
\end{equation}
where the $n\times n$ matrices $X^{(M)}=\left[ \langle \Psi _i\mid \Psi
_j\rangle ^M\right] $, $\Gamma =diag(\gamma _1,\gamma _2,...\gamma _n)$, $%
C=\left[ c_{ij}\right] $, $T=\left[ t_{ij}\right] $, we call $\Gamma $ as
correctly discriminating probabilistic matrix and $T$ as error
discriminating probabilistic matrix. $CC^{+}$ is semipositive definite,that
yield

\begin{equation}
X^{(M)}-(\sqrt{\Gamma }+T\ )(\sqrt{\Gamma }+T\ ^{+})\geq 0,
\end{equation}
where $\geq 0$ means semipositive definite. This inequality give a general
bound among the initial states matrix $X$, correctly discriminating
probabilistic matrix $\Gamma $ and error discriminating probabilistic matrix 
$T$. We don't make constraint on the initial states, which means we don't
demand that the initial states must be linearly independent. In the
following we discuss Inequality.(4) and find many former works about
discrimination can be derived from this inequality.

Denote $B=\sqrt{\Gamma }+T\ $,we rewrite inequality.(4) as

\begin{equation}
X^{(M)}-BB^{+}\geq 0.
\end{equation}

We begin our discuss with giving a condition that yield $\left\{ \left| \Psi
_i\right\rangle ^{\otimes M},i=1,2,...,n\right\} $ are linearly independent,
that is

\begin{equation}
\sqrt{\gamma _i}>\sum_jt_{ij},
\end{equation}
where $i=1,2,...,n$. Condition (6) yield $X^{(M)}$ is a positive definite
matrix [19], which means$\left\{ \left| \Psi _i\right\rangle ^{\otimes
M},i=1,2,...,n\right\} $ are linearly independent. We can derive from
inequality.(6) that $P_D^{(i)}>P_E^{(i)}$, which means the correctly
discriminating probability is great than that of error. We consider a
special situation that $T=0$. To obtain maximum of correctly discriminating
probability $\gamma _i>0$, the inequality (6) must be satisfied and $\left\{
\left| \Psi _i\right\rangle ^{\otimes M},i=1,2,...,n\right\} $ must be
linearly independent, which means only linearly independent states can be
discriminated with non-zero probability if we demand there are no error
existence. Such result has been obtained by Duan and Guo [11] and Chefles
and Barnett [6]. We can also give the maximum discriminating probability
that is determined by such inequality

\begin{equation}
X^{(M)}-\Gamma \geq 0.
\end{equation}
\quad

This inequality is just a generalization of the optimal efficiencies for
exactly probabilistically discriminating which has been given by Duan and
Guo [11] and we also obtain such inequality in [20].

In the following discussion we are only concerned with the discrimination of
two states. This situation is the most important and many valuable works has
been done to this. We will give the most general bound between the correctly
discriminating probability and that of error in this situation using
inequality.(4) and this bound comes to the former results in different
special conditions.

Consider a quantum system prepared in one of the two states $\left| \psi
_{\pm }\right\rangle ^{\otimes M}$ .We are not told which of the states the
system is in, although we do know that it has some probability of being in
either. Denote $P_{IP}=\langle \psi _{+}\mid \psi _{-}\rangle ^M$.We
represent $X$, $\Gamma $ and $T$ as $X=\left( 
\begin{array}{cc}
1 & P_{IP} \\ 
P_{IP}^{*} & 1
\end{array}
\right) $, $\Gamma =diag\left( \sqrt{P_D^{+}},\sqrt{P_D^{-}}\right) $, $%
T=\left( 
\begin{array}{cc}
0 & \sqrt{P_E^{+}} \\ 
\sqrt{P_E^{-}} & 0
\end{array}
\right) $ and inequality.(4) yield

\begin{equation}
(1-P_D^{+}-P_E^{+})(1-P_D^{-}-P_E^{-})\geq (P_{IP}-\sqrt{P_D^{+}P_E^{-}}-%
\sqrt{P_D^{-}P_E^{+}})(P_{IP}^{*}-\sqrt{P_D^{+}P_E^{-}}-\sqrt{P_D^{-}P_E^{+}}%
).
\end{equation}

For $P_I^{+}=1-P_D^{+}-P_E^{+}$, $P_I^{-}=1-P_D^{-}-P_E^{-}$, we can rewrite
inequality.(8) as

\begin{equation}
P_I^{+}P_I^{-}\geq \left| P_{IP}-\sqrt{P_D^{+}P_E^{-}}-\sqrt{P_D^{-}P_E^{+}}%
\right| ^2.
\end{equation}

This inequality is just the most general bound among the discriminating
probabilities of correct ($P_D^{+}$, $P_D^{-}$),error ($P_E^{+}$, $P_E^{-}$)
and inconclusive ($P_I^{+}$, $P_I^{-}$). In the following we will give some
special conditions and simple inequality.(9).

{\bf 1. }We let $P_I^{+}=P_I^{-}=0$, which means that we don't give
inconclusive results, then inequality.(9) yield

\begin{equation}
P_{IP}=\sqrt{P_D^{+}(1-P_D^{-})}+\sqrt{P_D^{-}(1-P_D^{+})}.
\end{equation}

Eq.(10) give a bound of the maximum correctly discriminating probabilities $%
P_D^{+}$, $P_D^{-}$ of the two states $\left| \psi _{+}\right\rangle $ and $%
\left| \psi _{-}\right\rangle $ given $M$ initial copies of each states. We
find $P_{IP}$ must be real, which means if the inter-produce of two states $%
\left| \psi _{+}\right\rangle $ and $\left| \psi _{-}\right\rangle $ is not
real we can't execute discrimination without inconclusive results.
Furthermore we can suppose $P_D^{+}=P_D^{-}=P_D$, and derive

\begin{equation}
P_D=\frac 12(1+\sqrt{1-P_{IP}^2}).
\end{equation}

When $M=1$, Eq.(11) just give the well-known Helstrom limit [2]

\begin{equation}
P_D=P_H=\frac 12(1+\sqrt{1-\left| \langle \psi _{+}\mid \psi _{-}\rangle
\right| ^2}).
\end{equation}

Eq.(11) give the absolute maximum probability of discriminating between two
states $\left| \psi _{\pm }\right\rangle $ with given $M$ initial copies.
The measurement it represents does not give inconclusive results, but will
incorrectly identify the states with probability $1-P_D$.

{\bf 2. }We suppose $P_{IP}=P_{IP}^{*}$, $P_D^{+}=P_D^{-}=P_D$, $%
P_E^{+}=P_E^{-}=P_E$, so $P_I^{+}=P_I^{-}=P_I$ and Eq.(9) yield

\begin{equation}
\frac 12(P_{IP}-P_I)\leq \sqrt{P_EP_D}\leq \frac 12(P_{IP}+P_I)\text{ .}
\end{equation}

Inequality (13) give a general lower bound on the combination of errors and
inconclusive result and corresponds to a family of measurements which
optimally interpolates between the Helstrom and IDP limits. If $P_{IP}\geq
P_I$ ,we can get

\begin{equation}
\frac 14(P_{IP}-P_I)^2\leq P_EP_D\leq \frac 14(P_{IP}+P_I)^2\text{ .}
\end{equation}

When $M=1$, the left side of inequality (14) is just the inequality that has
been obtained by Chefles and Barnett [8]. If $P_{IP}<P_I$, we have 
\begin{equation}
P_EP_D\leq \frac 14(P_{IP}+P_I)^2
\end{equation}

{\bf 3. }We suppose $P_E^{+}=P_E^{-}=0$, and inequality (8) yield 
\begin{equation}
(1-P_D^{+})(1-P_D^{-})\geq \left| P_{IP}\right| ^2.
\end{equation}

So we obtain 
\begin{equation}
\frac{P_D^{+}+P_D^{-}}2\leq 1-\left| \langle \psi _{+}\mid \psi _{-}\rangle
\right| ^M
\end{equation}

When $P_D^{+}=P_D^{-}=P_{IDP}$, we give $P_{IDP}=1-\left| \langle \psi
_{+}\mid \psi _{-}\rangle \right| ^M=1-P_{IP}$, which is a generalization of
IDP limit.

Denote $P_S=P_D+P_E$. It is obvious that $P_S$ is the probability with which
we measurement the probe system and get $\left| P_0\right\rangle $ after the
unitary evolution in Eq.(2). Comparing with [9], we can say $P_S$ is just
the probability that two states can be separated by an arbitrary degree with
states separating operation. In similar way we can denote $%
P_S^{(i)}=P_D^{(i)}+P_E^{(i)}$ and find the unitary evolution $\hat U$ in
Eq.(2) just transfer states $\left| \Psi _i\right\rangle _A^{\otimes
M}\left| P_0\right\rangle $ into states $\frac 1{\sqrt{P_S^{(i)}}}\left( 
\sqrt{\gamma _i}\left| \varphi _i\right\rangle _A+\sum_jt_{ij}\left| \varphi
_j\right\rangle _A\right) \left| P_0\right\rangle $ with transfer
probability $P_S^{(i)}$ and realize states separating operate. We can give $%
P_D=P_SP_H$, and $P_E=P_S(1-P_H)$, where $P_H$ is the Helstrom limit for $M$
initial states. With these representations, inequality (13) yield

\begin{equation}
P_E\geq \frac 12\left( P_S-\sqrt{P_S^2-(P_S-P_{IDP})^2}\right) \text{\quad
if }P_S\geq P_{IDP}
\end{equation}

This inequality is just that has been obtained by Chefles and Barnett [9]
when $M=1$, which give a bound on the error probability $P_E$ given a fixed
value of the probability $P_S$ and it is equivalent to inequality (15).

So far we have constructed general discrimination of the states secretly
chosen from a certain set $S$ given $M$ initial copies of each state by a
combination of unitary evolution together with measurements. We have derived
the matrix inequality, which describe the bound among three different
discriminating results: correct, error and inconclusive. For different $n$%
-state,we give a condition which yield such $n$-state are linearly
independent and find the result of Duan and Guo is just obtained in a
special situation of our condition. For two states, we find the most general
bound (inequality (9)) among the discriminating probabilities of correctly ($%
P_D^{+}$, $P_D^{-}$),error ($P_E^{+}$,$P_E^{-}$) and inconclusive ($P_I^{+}$,%
$P_I^{-}$) and the former works are the results of different applications of
our bound in different situation.

This work was supported by the National Natural Science Foundation of China.%
\newpage\


\begin{references}
\bibitem{}  See reference [11], Duan and Guo give a Lemma that if two sets
of states $\left| \phi _1\right\rangle ,\left| \phi _2\right\rangle
,...,\left| \phi _n\right\rangle ,$ and $\left| \widetilde{\phi }%
_1\right\rangle ,\left| \widetilde{\phi }_2\right\rangle ,...,\left| 
\widetilde{\phi }_n\right\rangle $ satisfy the condition $\langle \phi
_i\mid \phi _j\rangle =\langle \widetilde{\phi }_i\mid \widetilde{\phi }%
_j\rangle ,$there exists a unitary operate $U$ to make $U\left| \phi
_i\right\rangle =\left| \widetilde{\phi }_i\right\rangle $,($i=1,2,...,n$).
\bibitem{}  Matrix $B$ satisfy condition (6). We can prove that the
determinant of such matrix is non-zero,which means $BB^{+}$ is positive
definite. So $X^{(M)\text{ }}$must be positive definite to satisfy
inequality (5).
\end{references}
\end{document}